\documentstyle[12pt]{article}
\def\beq{\begin{equation}}
\def\eeq{\end{equation}}

\begin{document}

\begin{titlepage}
\begin{flushright}
BA-99-69\\
\end{flushright}

\begin{center}
{\Large\bf   Flavor Problem, Proton Decay And Neutrino \\
Oscillations
In SUSY Models With Anomalous ${\cal U}(1)$}

\end{center}
\vspace{0.5cm}
\begin{center}
{\large Qaisar Shafi$^{a}$ {}~and
{}~Zurab Tavartkiladze$^{b}$ }
\vspace{0.5cm}

$^a${\em Bartol Research Institute, University of Delaware,
Newark, DE 19716, USA \\

$^b$ Institute of Physics, Georgian Academy of Sciences,
380077 Tbilisi, Georgia}\\
\end{center}

\vspace{1.0cm}

\begin{abstract}

We discuss how realistic supersymmetric
models can be constructed by employing an anomalous ${\cal U}(1)$
flavor symmetry which also mediates supersymmetry breaking.
A judicious choice of ${\cal U}(1)$ charges enables the first two
squark families to be sufficiently heavy 
($\stackrel{>}{_\sim }10$~TeV), so that flavor changing neutral 
currents as well as dimension five nucleon decay
are adequately suppressed. 
Using the $SU(5)$ example, the charged fermion mass hierarchies, 
magnitudes of the CKM matrix elements, as well as the 
observed neutrino 
oscillations are simultaneously accommodated. 
We estimate the proton lifetime to be 
$\tau_p\sim 10^3\cdot \tau_p[{\rm minimal}~SU(5)]$, with the decay mode
$p\to K\mu $ being comparable to $p\to K\nu_{\mu, \tau}$.

\end{abstract}

\end{titlepage}

\section{Introduction}


Notwithstanding their enormous theoretical appeal,
supersymmetric (SUSY) models provide several important challenges to the 
model builder. These include the problem of flavor changing neutral currents
(FCNC), dimension five (including Planck scale suppressed) proton decay, and
CP violating phases.
Flavor non-conservation in 
SUSY theories is referred to as the supersymmetric flavor problem
and is closely tied with the mechanism of SUSY breaking. 
New sources for FCNC in SUSY theories can arise from non-universal  
sparticle soft masses, and from a non-alignment 
(non-proportionality) 
of trilinear soft terms with the charged fermion Yukawa matrices.
In $N=1$ minimal supergravity (SUGRA) \cite{barb}, universality and 
proportionality holds at the Planck scale ($M_P=2.4\cdot 10^{18}$~GeV).
For estimating flavor changing processes one should 
renormalize the soft SUSY 
breaking terms between $M_P$ and the SUSY breaking/electroweak scales. If 
a GUT 
scenario is considered one should also integrate out the heavy states
which 
decouple at $M_G$, the GUT scale. These two procedures violate 
universality and proportionality \cite{pom}, which could
cause problems with FCNC. In gauge 
mediated SUSY breaking 
alignment holds at the low energy scale and FCNC are 
adequately suppressed.

An alternative approach for resolving the supersymmetric flavor problem
is the so called decoupling solution \cite{dec, nelson, nomura}, 
in which the FCNC are 
suppressed by 
large squark and slepton masses. In order to satisfy the existing
experimental bounds
\cite{bounds} it is sufficient to have squarks (sleptons) in the 
mass range $\stackrel{>}{_\sim }10$~TeV. On the other hand, to avoid 
spoiling the gauge hierarchy, the stop mass 
(and also $\tilde{b}, \tilde{\tau }$ in case of large $\tan \beta $) 
should not exceed $\sim 1$~TeV.
The sparticles corresponding to the first two generations can be 
heavier, since their interactions with the higgs fields are suppressed 
by their Yukawa couplings.  
In the charged fermion sector we have, of course, the opposite 
hierarchical picture! A priori, without any symmetry reasons, it seems 
quite surprising to have a mass spectrum with such an inverse hierarchy.

In a recently proposed scenario \cite{ano}, one possible
mediator of SUSY breaking was assumed to be an anomalous ${\cal U}(1)$
symmetry,
so that SUSY breaking mainly occurs through a non-zero 
$D_A$-term (of ${\cal U}(1)$), while the contributions
from $F$-terms 
are relatively
suppressed. Sparticles will gain soft masses if their 
${\cal U}(1)$ charge is non zero. Otherwise, their soft
mass will be relatively suppressed. It is tempting to exploit this
${\cal U}(1)$ also as a flavor symmetry \cite{anu1}
(for the neutrino oscillation scenarios with ${\cal U}(1)$
flavor symmetry within MSSM and various GUTs see \cite{nuu1}, 
\cite{maxmix}-\cite{bimaxso10}).
Since the top quark mass is close to the electroweak symmetry breaking 
scale, it is natural to assume that it arises through a 
renormalizable Yukawa coupling 
The ${\cal U}(1)$ charges of the higgs superfields are  
taken to be zero. Note that in the absence of additional 
symmetries the $\mu $ problem remains unresolved.
The Yukawa couplings of the light families can be 
suppressed by prescribing them appropriate ${\cal U}(1)$ charges. 
It follows that sparticles corresponding to the light fermions will
have large soft masses in comparison 
to their counterparts from the third family. If the 
contribution to the soft masses from $D_A$-term is dominant and in the 
$10$~TeV range, the supersymmetric flavor problem will be resolved.

In this paper we attempt to develop this approach within the framework of 
$SU(5)$ GUT and study some of its phenomenological implications
(for earlier related works see \cite{nomura}-\cite{nima}). Employing an 
anomalous ${\cal U}(1)$ as a mediator of SUSY breaking, and as a flavor 
symmetry, we obtain a suitable
mass spectrum for proper suppression of FCNC. It turns out
that this also leads to a strong (and desirable) suppression of 
the dominant nucleon decay in minimal SUSY $SU(5)$, since 
in the internal loops of the $d=6$ nucleon decay diagrams, there 
appear sparticles belonging to the first two families.
In our scenario the dominant decays occur through 
diagrams in which sparticles of the third generation participate, and for 
adequate suppression the regime with intermediate or low $\tan \beta $
is required. It is worth stressing that the neutrino 
and charged lepton decay channels are comparable in magnitude, with the 
proton lifetime estimated 
to be $\tau_p\sim 10^3\tau_0$ (where 
$\tau_0\sim 10^{29\pm 2}$~yr. is the proton lifetime in  
minimal SUSY $SU(5)$, assuming squark and gaugino masses around 
$1$~TeV). Due to ${\cal U}(1)$ flavor symmetry, all Planck scale mediated
$d=5$ baryon number violating operators are also adequately suppressed. 
The model is also compatible with the various neutrino oscillation 
scenarios that are in agreement with the atmospheric and solar 
neutrino data 
\cite{atmSK, solSK}. We stress bi-maximal vacuum neutrino mixing scenario, 
with 
the ${\cal U}(1)$ symmetry once again playing a crucial role 
\cite{maxmix, bimaxsu5, bimaxso10}.
We also indicate how the large and small mixing angle MSW oscillations 
for resolving the solar neutrino anomaly can be realized. 

The paper is organized as follows: in Section 2 we discuss SUSY breaking 
through an anomalous ${\cal U}(1)$ symmetry, and show how the desirable 
sparticle spectrum needed for suppression of FCNC can be obtained. Some 
necessary conditions which should be satisfied are pointed out. We also
discuss suppression of nucleon decay and present the appropriate suppression 
factors which do not depend on GUT physics, but are closely tied 
to the low energy sector. In Section 3 we present an $SU(5)$ example in 
which (the same) anomalous ${\cal U}(1)$ symmetry is exploited as a flavor 
symmetry to provide a
natural understanding of hierarchies between charged fermion
masses and their mixings. 
We briefly explain how the bad asymptotic $SU(5)$ mass 
relations $\hat{M}_d^0=\hat{M}_e^0$ involving the
light families are avoided in our 
approach. We discuss the various neutrino oscillation 
scenarios which 
simultaneously accommodate the atmospheric and solar neutrino puzzles.  
Estimates for the nucleon decay widths are also presented.
Our conclusions are summarized in Section 4.

\section{SUSY breaking anomalous ${\cal U}(1)$ , FCNC and nucleon decay}

We employ the proposal of ref. \cite{ano} and consider an
anomalous ${\cal U}(1)$ symmetry as a mediator of SUSY breaking.
It is well known that  
anomalous ${\cal U}(1)$ symmetries often emerges from strings. The
cancellation of its anomalies occur through the 
Green-Schwarz mechanism \cite{green}, and the
associated Fayet-Iliopoulos term is given by \cite{fi}

\beq
\xi \int d^4\theta V_A~,~~~~
\xi =\frac{g_A^2M_P^2}{192\pi^2}{\rm Tr}Q~.
\label{fi}
\eeq
The $D_A$-term is
\begin{equation}
\frac{g_A^2}{8}D_A^2=\frac{g_A^2}{8}
\left(\Sigma Q_a|\varphi_a |^2+\xi \right)^2~,
\label{da}
\end{equation}
where $Q_a$ is the `anomalous' charge of $\varphi_a $ superfield.

Let us introduce a singlet superfield $X$ with ${\cal U}(1)$
charge $Q_{X}=-1$.
Assuming ${\rm Tr}Q>0$ ($\xi >0$), the cancellation of (\ref{da})
fixes the VEV of the scalar component of $X$:
\beq
\langle X\rangle =\sqrt{\xi }~,
\label{vevx}
\eeq
with SUSY unbroken at this stage. Including a mass term for $X$
in the superpotential,

\beq
W_m=\frac{m}{2}X^2~,
\label{massx}
\eeq
the cancellation of $D_A$ will be partial, and SUSY will be broken
due to non-zero $F$ and $D$ terms. Taking into account 
(\ref{da}) and (\ref{massx}), the potential for $X$ will have the form

\beq
V=m^2|X|^2+\frac{g_A^2}{8}
\left(\xi -|X|^2\right)^2~.
\label{potx}
\eeq
Minimization of (\ref{potx}) gives

\beq
X^2=\xi -\frac{4m^2}{g_A^2}~,
\label{solx}
\eeq
along which

\beq
\langle D_A\rangle =\frac{4m^2}{g_A^2}~,~~~~
\langle F_X\rangle \simeq m\sqrt{\xi }~.
\label{fdx}
\eeq
From (\ref{da}), taking into account (\ref{solx}), (\ref{fdx}),
for the soft scalar masses squared (${\rm mass}^2$) we have

\beq
m^2_{\tilde{\varphi }_a}=Q_am^2~.
\label{masssc}
\eeq
Thus, the scalar components of superfields which have non-zero
${\cal U}(1)$ charges gain masses through $\langle D_A\rangle$.

We will assume that the VEV of $X$ is somewhat below $M_P$,
namely

\beq
\frac{\langle X\rangle }{M_P}\equiv \epsilon \simeq 0.22~,
\label{epsx}
\eeq
while the scale $m$ is in the range $\sim 10$~TeV (see below). 
Those states
which have zero ${\cal U}(1)$ charges will gain soft masses
of the order of gravitino mass $m_{3/2}$ from the K\"ahler potential 

\beq
m_{3/2}=\frac{F_X}{\sqrt{3}M_P}=m\frac{\epsilon }{\sqrt{3}}~,
\label{gravmass}
\eeq
which, for $m=10$~TeV, is relatively
suppressed ($\sim 1$~TeV). 
The gaugino masses also will have the same magnitudes

\beq
M_{\tilde{G}_i}\sim m_{3/2}\sim 1~{\rm TeV}~.
\label{gaugmass}
\eeq

The mass term (\ref{massx}) violates the ${\cal U}(1)$ symmetry
and is taken to be in the $10$~TeV range. Its origin 
may lie in a strong
dynamics where $m$ is replaced by the VEV of some superfield(s) 
\cite{ano, gia}.
One possibility is to introduce a singlet superfield $Z$ with $Q_Z=2$,
and vector-like superfields $\bar Q+Q$ ($Q_{\bar Q}=Q_Q=0$),  
assumed to be 
a doublet-antidoublet pair of a strong $SU(2)$ gauge group. Then, 
imposing an additional global symmetry,

\beq
Z\to e^{{\rm i}\alpha }Z~,~~~~
\bar QQ\to e^{-{\rm i}\alpha }\bar QQ~,
\label{glob}
\eeq
the lowest term in the superpotential will be

\beq
W_0=\lambda \frac{\bar QQ}{M_P^2}ZX^2~.
\label{lowest}
\eeq
Assuming that $SU(2)$ becomes strong at scale $\Lambda $, the
non-perturbative superpotential induced by the instantons will have the 
form \cite{inst}

\beq
W_{{\rm inst}}=\frac{\Lambda ^5}{\bar QQ}~,
\label{inst}
\eeq
and the scalar superpotential will be
\footnote{The non-perturbative term (\ref{inst}) violates 
global symmetry in (\ref{glob}). This can happen if 
the symmetry is anomalous.}:

\beq
W_s=\lambda \frac{\bar QQ}{M_P^2}ZX^2+
\frac{\Lambda ^5}{\bar QQ}~.
\label{sup}
\eeq

The potential built from the $F$ and $D$-terms has the form
\beq
V_s=\Sigma |F_a|^2+\frac{g_A^2}{8}D_A^2~,
\label{pot1}
\eeq
where

\beq
F_a=\frac{d W_s}{d \varphi_a}~,~~~~~~~~D_A=\xi -|X|^2+2|Z|^2~.
\label{dxz}
\eeq
It is easy to verify that there is no solution along which the $F$ and
$D$-terms simultaneously vanish. Minimization of (\ref{pot1}) gives
the following solutions
 
$$
X^2=\frac{4}{3}\xi -\frac{16}{3}\frac{m^2}{g_A^2}~,~~~~
Z^2=\frac{1}{6}\xi -\frac{2}{3}\frac{m^2}{g_A^2}~,
$$
\beq
\bar Q^4=Q^4=\frac{9m^2M_P^2}{2\lambda}
\left(1-\frac{9\sqrt{3}}{8}\frac{mM_P^2}{\xi \sqrt{\xi }}\right)~,
\label{sol}
\eeq
where

\beq
m^2=\frac{\sqrt{6}\Lambda^5}{\xi \sqrt{\xi }}~.
\label{masgen}
\eeq
From (\ref{epsx}), (\ref{sol}) we find
\beq
\epsilon_Z\equiv \frac{\langle Z\rangle}{M_P}=
\frac{1}{2\sqrt{2}}\epsilon \simeq 0.08~,~~~~
\sqrt{\xi }=\frac{\sqrt{3}}{2}M_P\epsilon~.
\label{epsz}
\eeq

Substituting (\ref{sol}) in (\ref{dxz}), we readily obtain the 
expression for $\langle D_A\rangle $ given in (\ref{fdx}), and 
expression (\ref{masssc}) (for calculating soft masses)
is valid. Assuming that 
$\Lambda \simeq 3.3\cdot 10^{12}$~GeV, from (\ref{masgen})
we obtain the desirable value for $m$($\simeq 10$~TeV). In this  
example, among the non-zero $F$-terms, it is $F_Z$ which dominates 
and provides the 
dominant contribution to the gravitino mass ($\sim 1$~TeV)
in (\ref{gravmass}).

Turning now to the question of FCNC, we require that the
`light' quark-lepton 
superfields carry non-zero 
${\cal U}(1)$ charges. This means that the soft masses of their scalar 
components are in the $10$~TeV range, which automatically 
suppresses  flavor 
changing processes  such as $K^0-\bar K^0$, $\mu \to e\gamma$
etc., thereby satisfying the present experimental bounds \cite{bounds}.
To prevent upsetting the gauge hierarchy, the third generation
up squarks must have masses no larger than a TeV or so \cite{nima}
(hence they have zero ${\cal U}(1)$ charge). The same applies to  
sbottom and stau for large $\tan \beta $
since, for $\lambda_b\sim \lambda_{\tau }\sim 1$, 
large masses ($\stackrel{>}{_\sim }10$~TeV) of $\tilde{b}$
and $\tilde{\tau }$ would spoil the gauge hierarchy.  

Although the tree level mass of the stop can be arranged to be in the 
$1$~TeV range by the ${\cal U}(1)$ symmetry, the $2$-loop contributions
from heavy sparticles of the first two generations can 
drive the stop ${\rm mass}^2$  
negative \cite{nima}. This is clearly unacceptable, and one proposal 
for avoiding it \cite{nomura}
requires the existence of new states in the multi-TeV range.
The dangerous contribution which comes from $2$-loop
diagrams is proportional to

\beq
\Sigma m_{\tilde{\varphi }_a}^2T_a~,
\label{2loop}
\eeq
where $m_{\tilde{\varphi }_a}^2$ (see (\ref{masssc})) is 
the soft ${\rm mass}^2$ of
$\tilde{\varphi}_a$, and $T_a$ is the Dynkin index of the appropriate 
representation. The 
representations and ${\cal U}(1)$ charges of the new states 
should be chosen so that (\ref{2loop}) vanishes,
namely

\beq
\Sigma Q_aT_a=0~.
\label{cond}
\eeq
We will later see how this is implemented in the $SU(5)$ example.

Let us now turn to some implications for proton decay.
We assume that $d=5$ baryon number 
violating operators arise from the couplings

\beq
qAqT+qBl\bar T~,
\label{qqt}
\eeq
after integration of color triplets $T, \bar T$ with mass 
$M_T\sim 2\cdot 10^{16}$~GeV (we consider triplet 
couplings with left-handed matter, which
provide the dominant contribution to nucleon decay).
After wino dressing of appropriate $d=5$ operators, the resulting $d=6$ 
operators causing proton to decay into 
the neutrino and charged lepton channels have the respective forms
\cite{his, 3rd}:

\beq
\frac{g_2^2}{M_T}
\alpha (u_a d^i_b)(d^j_c\nu^k) 
\varepsilon^{abc}~,
\label{d6nu}
\eeq
 
\beq
\frac{g_2^2}{M_T}
\alpha' (u_a d^i_b)
(u_ce^j)
\varepsilon^{abc}~,
\label{d6e}
\eeq
where

$$
\alpha=
-[(L_d^{+}\hat{B}L_e)_{jk}
(L_u^{+}\hat{A}L_d^{*})_{mn}+
(L_d^{+}\hat{A}L_u^{*})_{jm}
(L_d^{+}\hat{B}L_e)_{nk}]
V_{mi}(V^{+})_{n1}
I(\tilde{u}^m,\tilde{d}^n)+ 
$$
\beq
[(L_u^{+}\hat{A}L_d^{*})_{1i}
(L_u^+\hat{B}L_e)_{mk}
-(L_d^{+}\hat{A}L_u^{*})_{im}
(L_u^{+}\hat{B}L_e)_{ik}]V_{mj}
I(\tilde{u^m},\tilde{e^k})~, 
\label{d6nu1}
\eeq

$$
\alpha' =
[-(L_u^{+}\hat{A}L_d^{*})_{1i}
(L_d^{+}\hat{B}L_e)_{mj}
+(L_u^{+}\hat{A}L_d^{*})_{1m}
(L_d^{+}\hat{B}L_e)_{ij}](V^{+})_{m1}
I(\tilde{d^m},\tilde{\nu^j})+ 
$$
\beq
[(L_u^{+}\hat{B}L_e)_{1j}
(L_u^{+}\hat{A}L_d^{*})_{mn}+
(L_u^{+}\hat{A}L_d^{*})_{1m}
(L_e^{T}\hat{B}^TL_u^{*})_{jn}]
(V^{+})_{m1}
V_{ni}
I(\tilde{d}^m, \tilde{u}^n)~. 
\label{d6e1}
\eeq
$L$'s are unitary matrices which rotate the left handed fermion states
to diagonalize the mass matrices, and $I$'s 
are functions obtained after loop integration and depend on the SUSY 
particle masses circulating inside the loop. For example \cite{his},

\beq
I(\tilde{u}, \tilde{d})=\frac {1}{16\pi^2}\frac
{m_{\tilde{W}}}{m_{\tilde{u}}^2- m_{\tilde{d}}^2} 
\left ( \frac{m_{\tilde{u}}^2}{m_{\tilde{u}}^2- 
m_{\tilde{W}}^2}\ln \frac{m_{\tilde{u}}^2}{m_{\tilde{W}}^2}- 
\frac{m_{\tilde{d}}^2}{m_{\tilde{d}}^2- 
m_{\tilde{W}}^2}\ln \frac{m_{\tilde{d}}^2}{m_{\tilde{W}}^2} \right )~,
\label{int}
\eeq
with similar expressions for $I(\tilde{d}, \tilde{\nu} )$ and
$I(\tilde{u},\tilde{e} )$.

Consider those  
diagrams in which  sparticles of the first two 
families participate. Since their masses are large 
($\stackrel{>}{_\sim }10$~TeV) compared 
to the case with minimal $N=1$ SUGRA, we expect considerably  
suppression of proton decay.  
For minimal $N=1$ SUGRA, 
$m_{\tilde{u}}\sim m_{\tilde{d}}\sim m_{\tilde{W}}\sim m_{3/2}\sim 1$~TeV,
and (\ref{int}) can be approximated by

\beq
I_0\approx \frac{1}{16\pi^2}\frac{1}{m_{3/2}}~.
\label{int0}
\eeq 
In the ${\cal U}(1)$ mediated SUSY breaking scenario, 
expression  (\ref{int}) takes the  
form

\beq
I'\approx \frac{1}{16\pi^2}\frac{m_{\tilde{W}}}{m_{\tilde{q}}^2}
\equiv \eta I_0
\label{int1}
\eeq
The nucleon lifetime in this case will be enhanced by the factor
$\frac{1}{\eta^2}\sim 10^4$.

Of course, there exist diagrams in which one sparticle from the 
third and one from the `light' families participate. In this case,
(\ref{int}) takes the form

\beq
I''\approx \frac{1}{16\pi^2}\frac{2m_{\tilde{W}}}{m_{\tilde{q}}^2}
\ln \frac{m_{\tilde{q}}}{m_{\tilde{W}}}
\equiv \eta' I_0
\label{int2}
\eeq
and the corresponding proton lifetime will be
$\sim \frac{1}{\eta'^2}\sim 500$ times large. 

As pointed out in \cite{3rd} (within minimal $N=1$ SUGRA), 
the contribution 
from diagrams in which sparticles from the third 
generation participate could be comparable with those 
arising from the light 
generation sparticle exchange. Since minimal SUSY $SU(5)$ gives 
unacceptably fast proton decay with $\tau_0\sim 10^{29\pm 2}$~yr, 
care must exercised in realistic model building 
(the situation is exacerbated if $\tan \beta $ is large).
This problem is easily avoided in the anomalous ${\cal U}(1)$
mediated SUSY breaking scenario. Note that since the 
dominant contribution
comes from the second term of (\ref{d6nu1}), and 
in the internal lines of the 
appropriate nucleon decay diagram there necessarily runs one slepton 
state, 
even if the latter belongs to third family, it can have mass in the $10$~TeV 
range if $\tan \beta $
is either of intermediate ($\sim 10-15$) or low value
(this is required for preserving the 
desired gauge hierarchy). Thus, thanks to the anomalous ${\cal U}(1)$
symmetry, in addition to avoiding dangerous FCNC, one can also obtain 
adequate
suppression of nucleon decay. Interestingly, this disfavors the large 
$\tan \beta $ regime which could be a characteristic feature of this 
class of models!

\section{An $SU(5)$ example}

Let us now consider in detail a SUSY $SU(5)$ GUT and
show how things discussed in the previous section work out
in practice.

\subsection{Charged fermion masses and mixings}

We exploit the anomalous ${\cal U}(1)$ as a flavor symmetry \cite{anu1}
to help provide a natural understanding of the hierarchies 
between the charged 
fermion masses and their mixings. In these considerations the parameter 
$\epsilon\simeq 0.22$ (see (\ref{epsx})) plays an essential role.
The three families of matter in $(10+\bar 5)$ representations, and higgs 
superfields $\bar H(\bar 5)+H(5)$ 
\footnote{We assume the presence of $Z_2$ matter parity
which distinguishes the matter and higgs superfields and prevents 
rapid proton decay.}
have the 
following transformation properties under ${\cal U}(1)$:

$$
Q_{10_3}=0~,~~Q_{10_2}=2~,~~Q_{10_1}=3
$$
\beq
Q_{\bar 5_1}=2+n~,~~~Q_{\bar 5_2}=Q_{\bar 5_3}=n~,~~
Q_{\bar H}=Q_H=0~.
\label{charges}
\eeq
The couplings relevant for the generation of up type quark masses 
are given by

\begin{equation}
\begin{array}{ccc}
&  {\begin{array}{ccc}
\hspace{-5mm}~~10_1 & \,\,10_2  & \,\,10_3

\end{array}}\\ \vspace{2mm}
\begin{array}{c}
10_1 \\ 10_2 \\10_3
 \end{array}\!\!\!\!\! &{\left(\begin{array}{ccc}
\,\,\epsilon^6~~ &\,\,\epsilon^5~~ &
\,\,\epsilon^3
\\
\,\,\epsilon^5~~   &\,\,\epsilon^4~~  &
\,\,\epsilon^2
 \\
\,\,\epsilon^3~~ &\,\,\epsilon^2~~ &\,\,1
\end{array}\right)H }~,
\end{array}  \!\!  ~~~~~
\label{up}
\end{equation}
while those responsible for down quark and charged lepton 
masses are

\begin{equation}
\begin{array}{ccc}
 & {\begin{array}{ccc}
\hspace{-5mm}\bar 5_1~ & \,\,\bar 5_2 ~~ & \,\,\bar 5_3 ~~ 

\end{array}}\\ \vspace{2mm}
\begin{array}{c}
10_1 \\ 10_2 \\10_3
 \end{array}\!\!\!\!\! &{\left(\begin{array}{ccc}
\,\,\epsilon^5~~ &\,\,\epsilon^3~~ &
\,\,\epsilon^3
\\
\,\,\epsilon^4~~   &\,\,\epsilon^2~~  &
\,\,\epsilon^2
 \\
\,\,\epsilon^2~~ &\,\,1~~ &\,\,1
\end{array}\right)\epsilon^n \bar H }~.
\end{array}  \!\!  ~~~~~
\label{downe}
\end{equation}
Upon diagonalization of (\ref{up}), (\ref{downe}) we obtain

\beq
\lambda_t\sim 1~,~~~\lambda_u : \lambda_c :  \lambda_t \sim
\epsilon^6 : \epsilon^4 :1~.
\label{lambdaup}
\eeq
$$
\lambda_b\sim \epsilon^n~,~~
\lambda_d :\lambda_s :\lambda_b \sim
\epsilon^5:\epsilon^2 :1~,~~~
$$
\beq
\lambda_{\tau }\sim \epsilon^n~,~~
\lambda_e :\lambda_{\mu } :\lambda_{\tau } \sim
\epsilon^5:\epsilon^2 :1~,
\label{lambdas}
\eeq
where $n=0, 1, 2$ determines the value of $\tan \beta $,

\beq
\tan \beta \sim \epsilon^n\frac{m_t}{m_b}~.~~
\label{tanbeta}
\eeq
From (\ref{up}) and (\ref{downe}), we obtain
\beq
V_{us} \sim \epsilon \ , \ V_{cb} \sim \epsilon^2 \ ,\ V_{ub} \sim
\epsilon^3~.
\label{ckm}
\eeq
We see that the ${\cal U}(1)$ symmetry yields desirable 
hierarchies (\ref{lambdaup}), (\ref{lambdas}) of
charged fermion Yukawa couplings as well as the magnitudes of the CKM 
matrix elements (\ref{ckm}).

The reader will note, however,  that
(\ref{downe}) implies the asymptotic mass relations 
$\hat{M}_d^0=\hat{M}_e^0$, which are unacceptable for the two 
light families.
This is readily avoided through the mechanism suggested in 
\cite{bimaxsu5} by employing two pairs of 
$(\overline{15}+15)_{1,2}$ matter states. Namely, with 
${\cal U}(1)$ charges

\beq
Q_{15_1}=-Q_{\overline{15}_1}=3~,~~~~~~
Q_{15_2}=-Q_{\overline{15}_2}=2~,
\label{ch15}
\eeq
consider the couplings

\begin{equation}
\begin{array}{cc}
 & {\begin{array}{ccc}
10_1&\,\,10_2&\,\,10_3~~~
\end{array}}\\ \vspace{2mm}
\begin{array}{c}
\overline{15}_1\\ \overline{15}_2

\end{array}\!\!\!\!\! &{\left(\begin{array}{ccc}
\,\, 1~~&
\,\,  0~~ &\,\, 0
\\
\,\, \epsilon ~~ &\,\,1~~&\,\, 0~
\end{array}\right)\Sigma }~,
\end{array}  \!\!~
\begin{array}{cc}
& {\begin{array}{cc}
15_1&\,\,
15_2~~~~~
\end{array}}\\ \vspace{2mm}
\begin{array}{c}
\overline{15}_1 \\ \overline{15}_2

\end{array}\!\!\!\!\! &{\left(\begin{array}{ccc}
\,\, 1~~
 &\,\,0
\\
\,\, \epsilon~~
&\,\,1
\end{array}\right)M_{15}~,
}
\end{array}
\label{1015}
\end{equation}                
where $\Sigma $ is the scalar $24$-plet whose VEV breaks $SU(5)$ 
down to $SU(3)_c\times SU(2)_L\times U(1)_Y$.
For $M_{15}\sim \langle \Sigma \rangle $, we see that the `light'
$q_{1,2}$ states reside both in $10_{1,2}$ and $15_{1,2}$
states with similar `weights'. At the same time, the other light states 
from $10$-plets ($u^c$ and $e^c$) will not be affected because the 
$15$-plets do not contain fragments with the relevant quantum numbers. Thus,
the relations $m_s^0=m_{\mu }^0$ and $m_d^0=m_e^0$ are avoided, while 
$m_b^0=m_{\tau }^0$ still holds since the terms in (\ref{1015})
do not affect $10_3$.

As far as the sparticle spectrum is concerned, since the superfields
$10_3, \bar H, H$ have zero ${\cal U}(1)$ charges, the soft masses of 
their scalar components will be in the $1$~TeV range,

\beq
m_{\tilde{10}_3}\sim m_{\bar H}\sim m_{H}
\sim m_{3/2}=1~{\rm TeV}~,
\label{soft3h}
\eeq
while for $10_{1,2}$ and $\bar 5_1$ we have

\beq
m_{\tilde{10}_1}\sim m_{\tilde{10}_2}\sim m_{\tilde{\bar 5}_1}
\sim m\sim 10~{\rm TeV}~.
\label{soft12}
\eeq
The soft masses of the scalar fragments from $\bar 5_{2,3}$
depend on the value of $n$, and for $n\neq 0$, they also will be in the
$10$~TeV range, which is preferred for proton stability.

In order to satisfy condition (\ref{cond}) and avoid color instability in 
our model, we will introduce one pair of  
$\bar F(\bar 5)+F(5)$ supermultiplets with ${\cal U}(1)$
charges

\beq
Q_{\bar F}=Q_F=-\frac{1}{2}(17+3n)~,
\label{QF}
\eeq
and with the following transformation properties under the 
symmetry in (\ref{glob}),

\beq
\bar FF\to e^{-(10+n){\rm i}\alpha }\bar FF~.
\label{glob1}
\eeq
The superpotential coupling which generates mass term for these states 
is given by

\beq
W_F=M_P\left(\frac{Z}{M_P}\right)^{10+n}
\left(\frac{X}{M_P}\right)^{3-n}\bar FF~,
\label{opF}
\eeq
from which, after substituting the VEVs 
(\ref{epsx}) and (\ref{epsz}), we obtain

\beq
M_F=M_P\epsilon_Z^{10+n}\epsilon^{3-n}=\left\{ \begin{array}{lll}
200~{\rm TeV} & \mbox{if $n=0$} \\
100~{\rm TeV} & \mbox{if $n=1$} \\
40~{\rm TeV} & \mbox{if $n=2$}
\end{array}
\right.
\label{massF}
\eeq
Therefore, the masses of these additional states are 
considerable more than a TeV range. 
It is easy to verify that $M_F^2$ 
dominates over the negative soft ${\rm mass}^2$ 
($=-\frac{17+3n}{2}m^2\simeq -(30~{\rm TeV})^2$) for all 
possible values of $n$($=0, 1, 2$), so that color will be unbroken.
Furthermore, taking into account (\ref{charges}) and (\ref{QF}), it is 
easily  
checked that the condition (\ref{cond}) which prevents the stop quark
${\rm mass}^2$ from becoming negative is automatically satisfied.

Finally, since the ${\cal U}(1)$ charges of $15_{1,2}$ states are 
the same as those
of $10_{1,2}$'s, the soft ${\rm mass}^2$ terms for light
$\tilde{q}_{1,2}$ fragments are unchanged so that 
(\ref{cond}), with the choice of charges in (\ref{QF}), still holds.

\subsection{Neutrino oscillations}

We next demonstrate how the solar
and atmospheric neutrino data can be accommodated within the 
$SU(5)$ scheme. We stress the
bi-maximal vacuum oscillation scenario, but also point out how 
the small (or large) mixing
angle MSW oscillations can be realized. Indeed, the picture  
is similar to our previously considered $SU(5)$ \cite{bimaxsu5}
and $SO(10)$ \cite{bimaxso10} scenarios
\footnote{For other scenarios of neutrino oscillation  with
${\cal U}(1)$ flavor symmetry within MSSM and various GUTs see
\cite{nuu1}.}.

Since the states $\bar 5_2$ and $\bar 5_3$ have the same ${\cal U}(1)$ charge 
(see (\ref{charges})), we can expect naturally large $\nu_{\mu }-\nu_{\tau }$
mixing. This also can be seen from the texture in (\ref{downe}). 
Introducing an $SU(5)$ singlet right handed neutrino ${\cal N}_3$
with suitable mass, the state
`$\nu_3$' can acquire the mass relevant for the atmospheric
neutrino puzzle. At this stage the other two neutrino states are massless.

Large 
$\nu_e-\nu_{\mu, \tau}$ mixing can be obtained by invoking the mechanism 
suggested in
\cite{maxmix} which naturally yields `maximal' mixings between neutrino
flavors.
For this we need two additional $SU(5)$ singlet states
${\cal N}_1$, ${\cal N}_2$. Under ${\cal U}(1)$, the 
${\cal N}_i$ states carry charges:

\beq
Q_{{\cal N}_1}=-Q_{{\cal N}_2}=n+2~,~~~~Q_{{\cal N}_3}=0~.
\label{Nu1z2}
\eeq
The relevant couplings are 
\beq
W_{{\cal N}_3}= M_{{\cal N}_3}{\cal N}_3^2+
\epsilon^n (a \epsilon^2 \bar 5_1+b\bar 5_2+c\bar 5_3)H{\cal N}_3 ,
\label{N3}
\eeq
\begin{equation}
\begin{array}{cc}
 & {\begin{array}{cc}
{\cal N}_1~&\,\,{\cal N}_2~~~~~~
\end{array}}\\ \vspace{2mm}
\begin{array}{c}
\bar 5_1\\ \bar 5_2 \\ \bar 5_3

\end{array}\!\!\!\!\! &{\left(\begin{array}{ccc}
\,\, \epsilon^{2n+4}~~ &
\,\,  1
\\
\,\, \epsilon^{2n+2}~~ &\,\,0   
\\
\,\, \epsilon^{2n+2}~~ &\,\,0
\end{array}\right)H }~,
\end{array}  \!\!~~~
\begin{array}{cc} 
 & {\begin{array}{cc}   
{\cal N}_1~&\,\,
{\cal N}_2~~~~~
\end{array}}\\ \vspace{2mm} 
\begin{array}{c}   
{\cal N}_1 \\ {\cal N}_2

\end{array}\!\!\!\!\! &{\left(\begin{array}{ccc}
\,\, \epsilon^{2n+4}
 &\,\,~~~1
\\
\,\, 1
&\,\,~~~0
\end{array}\right)M_{{\cal N}}~,
}
\end{array}~~~
\label{Ns}
\end{equation}  
where $a, b, c$ are dimensionless coefficients.
Note that there also exists the coupling
$M'\epsilon^{2+n}{\cal N}_1{\cal N}_3$ which, if properly suppressed
(see below), will not be relevant. 

Let us choose the basis in which the charged lepton matrix (\ref{downe})
is diagonal. This choice is convenient because the matrix which
diagonalizes the neutrino mass matrix will then coincide 
with the lepton mixing
matrix. The hierarchical structure of the couplings in (\ref{N3})
will not be altered, while the `Dirac' and `Majorana' masses from (\ref{Ns})
will respectively have the forms
\begin{equation}
\begin{array}{cc}
m_D=\!\!\!\!\! &{\left(\begin{array}{ccc}
\,\, \epsilon^{2n+4}~ &
\,\,  1
\\
\,\, \epsilon^{2n+2}~ &\,\,\epsilon^2
\\
\,\, \epsilon^{2n+2}~ &\,\,\epsilon^2
\end{array}\right)h_u }~,
\end{array}
\begin{array}{cc}

~~M_R=\!\!\!\!\! &{\left(\begin{array}{ccc}
\,\, \epsilon^4
 &\,\,1
\\
\,\, 1
&\,\,0
\end{array}\right)M_{\cal N}~.
}
\end{array}
\label{dirmaj}
\end{equation}

Taking 
\beq
M'\ll M_{{\cal N}_3}/\epsilon^{2n}~,~~~~ 
M_{\cal N}\stackrel{>}{_\sim }\frac{M'^2\epsilon^{2n}}{M_{{\cal N}_3}}
\label{cond1}
\eeq
and the other coefficients of order unity, integration of the 
${\cal N}$ states leads to the following `light' neutrino mass matrix:
\beq
\hat{m}_{\nu }=\hat{A}m+\hat{B}m'~,
\label{matnu}
\eeq
where
\beq
m\equiv \frac{\epsilon^{2n}h_u^2}{M_{{\cal N}_3}}~,~~~~
m'\equiv\frac{\epsilon^{2n+2}h_u^2}{M_{\cal N}}~,
\label{scales}
\eeq
$$
\begin{array}{ccc}
 & {\begin{array}{ccc}
~& \,\,~  & \,\,~~
\end{array}}\\ \vspace{2mm}
\hat{A}=
\begin{array}{c}
\\  \\
 \end{array}\!\!\!\!\!\!\!\!&{\left(\begin{array}{ccc}
\,\,a^2\epsilon^4  &\,\,~~ab\epsilon^2 &
\,\,~~ac\epsilon^2
\\
\,\,ab\epsilon^2   &\,\,~~b^2  &
\,\,~~bc
 \\
\,\, ac\epsilon^2 &\,\,~~bc  &\,\,~~c^2
\end{array}\right)m }~, 
\end{array}  \!\!  ~~
$$
\beq
\begin{array}{ccc} 
 & {\begin{array}{ccc}  
~~\,\,~  & \,\,~~~~
\end{array}}\\ \vspace{2mm}
\hat{B}=
\begin{array}{c}
\\  \\
 \end{array}\!\!\!\!\!\!\!\!&{\left(\begin{array}{ccc}
\,\,\epsilon^2 &\,\,~1 &
\,\,~1
\\
\,\,1   &\,\,~\epsilon^2  &
\,\,~\epsilon^2
 \\
\,\, 1 &\,\,~ \epsilon^2  &\,\,~\epsilon^2
\end{array}\right)m' }
\end{array}  \!\!  ~.
\label{AB}
\end{equation}
For

\beq
M_{{\cal N}_3}\simeq \epsilon^{2n}\cdot 10^{15}~{\rm GeV}~,~~~~
M_{\cal N}\simeq \epsilon^{2n+2}\cdot 10^{18}~{\rm GeV}~,
\label{scales1}
\eeq
the `light' eigenvalues are
$$  
m_{\nu_3}\simeq m(b^2+c^2+a^2\epsilon^4)
\sim 3\cdot 10^{-2}~{\rm eV}~,
$$  
\beq
m_{\nu_1 }\simeq m_{\nu_2 }\simeq m'\sim 3\cdot 10^{-5}~{\rm eV}~.
\label{masses}
\eeq
Ignoring CP violation the neutrino mass matrix (\ref{matnu})
can be diagonalized by the
orthogonal transformation $\nu_{\alpha }=U_{\nu}^{\alpha i}\nu_i$, where
$\alpha =e, \mu, \tau $ denotes flavor indices, $i=1, 2, 3$
the mass eigenstates,
and $U_{\nu }$ takes the form
\beq
\begin{array}{ccc}
U_{\nu }=~~
\!\!\!\!\!\!\!\!&{\left(\begin{array}{ccc}
\,\,\frac{1}{\sqrt{2}} &\,\,~~\frac{1}{\sqrt{2}} &
\,\,~~s_1
\\
\,\,-\frac{1}{\sqrt{2}}c_{\theta }  &\,\,~~~~\frac{1}{\sqrt{2}}c_{\theta }
&\,\,~~s_{\theta }
\\
\,\, ~~\frac{1}{\sqrt{2}}s_{\theta } &\,\,~
-\frac{1}{\sqrt{2}}s_{\theta }  &\,\,~~c_{\theta }
\end{array}\right) }
\end{array}~,
\label{lepckm}
\end{equation}
with
\beq
\tan \theta =\frac{b}{c}~,~~~~s_1=\frac{a\epsilon^2}{\sqrt{b^2+c^2}}~,
\label{angles}
\eeq
and $s_{\theta }\equiv \sin \theta $, $c_{\theta }\equiv \cos \theta $.
From (\ref{matnu})-(\ref{angles}) the solar and atmospheric neutrino
oscillation
parameters are
$$
\Delta m^2_{21 }\sim 2m'^2\epsilon^2\simeq 10^{-10}~{\rm eV}^2~,
$$  
\beq
{\cal A}(\nu_e \to \nu_{\mu , \tau }) =1-{\cal O}(\epsilon^4)~,
\label{solosc}
\eeq
$$
\Delta m^2_{32}\simeq m_{\nu_3}^2\sim 10^{-3}~{\rm eV}^2~,
$$
\beq
{\cal A}(\nu_{\mu }\to \nu_{\tau })=\frac{4b^2c^2}{(b^2+c^2)^2}-
{\cal O}(\epsilon^4)~,
\label{atmosc}
\eeq
where the oscillation amplitudes are defined as
\beq
{\cal A}(\nu_{\alpha }\to \nu_{\beta })=
4\Sigma_{j<k}U_{\nu }^{\alpha j}U_{\nu }^{\alpha k}
U_{\nu }^{\beta j}U_{\nu }^{\beta k}~.
\label{defamp}
\eeq

We see that the solar neutrino puzzle is explained by maximal vacuum
oscillations of $\nu_e $ into $\nu_{\mu, \tau }$. For $b\sim c$ the
$\nu_{\mu }-\nu_{\tau }$ mixing is naturally large, as suggested by the
atmospheric anomaly. For $b\simeq c$ the $\nu_{\mu }-\nu_{\tau }$ mixing
will be even maximal, and $\nu_e $ oscillations will be $50\%$ into
$\nu_{\mu }$ and $50\%$ into $\nu_{\tau }$. 

As far as the small angle MSW solution for the solar neutrino puzzle is 
concerned, from
(\ref{downe})
we see that the expected mixing between $\nu_e$ and $\nu_{\mu, \tau }$
states is $\sim \epsilon^2 $, which provides the desirable
value $\sin^2 2\theta \sim 4\epsilon^4\simeq 10^{-2}$.
To obtain
$\nu_e-\nu_{\mu, \tau }$ oscillations, we can introduce a $SU(5)$ singlet
state $N$ (instead of ${\cal N}_{1, 2}$ states), which will provide mass 
in the $10^{-3}$~eV range to the
`$\nu_2$' state, so that the small angle MSW oscillation for explaining the
solar neutrino deficit is realized. 

Large mixing angle MSW solution is obtained by keeping the 
${\cal N}_{1, 2}$ states with the transformation properties in (\ref{Nu1z2}).
Maximal $\nu_e-\nu_{\mu, \tau }$ oscillations will still hold, and the 
desired scale ($\sim 10^{-6}~{\rm eV}^2$) can be generated by taking 
$M_{\cal N}\simeq \epsilon^{2n+2}\cdot 10^{16}$~GeV in (\ref{scales}).
The oscillation picture (\ref{atmosc})
for the atmospheric neutrinos will be unchanged.

\subsection{Nucleon decay in $SU(5)$}

Turning to the issue of nucleon decay in $SU(5)$ , we 
will take $n\neq 0$ in (\ref{charges}), which provides soft masses for 
$\bar 5_{2,3}$ states in the $10$~TeV range. As pointed out in 
section 2, this will enhance proton stability. For decays with 
neutrino emission, in the relevant diagrams there circulate 
$\tilde{t}$ and $\tilde{\mu }(\tilde{\tau })$. Using the forms of
(\ref{up}), (\ref{downe}), and taking into account (\ref{d6nu1}),
(\ref{int1}), one estimates from (\ref{d6nu}),

\beq
\tau (p\to K\nu_{\mu , \tau})\sim \frac{1}{\eta'^2}
\left(\frac{\sin^2 \theta_c}{V_{ts}}\right)^2\tau_0
\sim 2\cdot 10^3\tau_0~,
\label{pknu}
\eeq
where $\theta_c$ is the Cabibbo angle and $\tau_0$ is the proton 
lifetime in minimal $SU(5)$ model combined with SUGRA
[in obtaining (\ref{pknu}), we 
took into account that 
$\tau_0\propto (\lambda_s\lambda_c\sin^2 \theta )^2$].

As far as decays with emission of charged leptons are concerned, there 
are diagrams inside which circulate $\tilde{t}$, $\tilde{b}$ states, which 
are 
in the $1$~TeV range. It turns out that these diagrams provide the dominant 
contribution to proton decay. However, considerable
suppression relative to minimal $SU(5)$ still  
occurs due to the small mixings between the third and light generations, 
and also because the baryon-meson-charged lepton matrix element is 
relatively suppressed \cite{his}. From all this, taking into account
(\ref{d6e}), (\ref{d6e1}), (\ref{up}), (\ref{downe}), for the 
dominant decay we find

\beq
\tau (p\to K\mu )\sim 10\left(
\frac{\sin^2 \theta_c}{V_{ub}}\right)^2\tau_0
\sim 10^3\tau_0~.
\label{pkmu}
\eeq

In summary, the color triplet mediated proton decay modes are 
adequately suppressed and interestingly, the decays into the charged lepton 
and neutrino channels are comparable.

Before concluding, let us note that the Planck scale suppressed baryon 
number 
violating $d=5$ operator $\frac{1}{M_P}q_1q_1q_2l_{2,3}$, which could 
cause unacceptably fast proton decay, is also suppressed, 
since it emerges from the coupling

\beq
\frac{1}{M_P}\left(\frac{X}{M_P}\right)^{8+n}
10_1 10_1 10_2 \bar 5_{2,3}~,
\label{planckd5}
\eeq
with the suppression guaranteed by the ${\cal U}(1)$ symmetry.

\section{Conclusions}

In this paper we have discussed SUSY models
which are accompanied by an anomalous ${\cal U}(1)$ symmetry. 
If the latter 
mediates SUSY breaking,  crucial
suppression of FCNC as well as dimension five
proton decay can be achieved. If the 
same ${\cal U}(1)$ also acts as flavor symmetry, one can provide a natural
qualitative 
explanation of the hierarchies between the charged fermion masses and the 
values of CKM matrix elements.

An example based on $SU(5)$ is worked out in detail, with
neutrino oscillations also taken into account. 
The ${\cal U}(1)$ flavor symmetry also adequately suppresses
the Planck scale induced baryon number violating $d=5$ operators.
The mechanisms discussed in this paper can be extended 
to a variety of GUTs such as $SO(10)$ and $SU(5+N)$.

\vspace{0.3cm}

This work was supported in part by  DOE under Grant No. DE-FG02-91ER40626
and by NATO, contract number CRG-970149.


\bibliographystyle{unsrt}

\end{document}